\documentclass[preprint,12pt]{elsarticle}
\usepackage[margin=2cm]{geometry}
\counterwithin{figure}{section}
\usepackage{rotating}
\usepackage{wrapfig}
\usepackage[table,xcdraw]{xcolor}

\usepackage{float}
\usepackage{booktabs, ltablex, makecell}
\usepackage{caption}
\usepackage{xurl}
\usepackage{hyperref}

\usepackage[numbers]{natbib}

\journal{TBD}

\begin{document}

\begin{frontmatter}


\author[label1]{Tatyana Dergunova}
\affiliation[label1]{organization={School of Engineering, Institute for Energy Systems, University of Edinburgh},
            addressline={Colin Maclaurin Road}, 
            city={Edinburg},
            postcode={EH9 3DW}, 
            country={UK}}

\title{Great Britain’s Hydrogen Infrastructure Development – Investment Priorities and Locational Flexibility}

\author[label1]{Andrew Lyden\corref{cor1}}
\ead{andrew.lyden@ed.ac.uk}
\cortext[cor1]{Corresponding author}



\begin{abstract}
Future pathways for Great Britain’s energy system decarbonisation have highlighted the importance of low-carbon hydrogen as an energy carrier and demand flexibility support. However, the potential application within various sectors (heating, industry, transport) and production capacity through different technologies (methane reformation with carbon capture, biomass gasification, electrolysis) is highly varying, introducing substantial uncertainties for hydrogen infrastructure development. This study sets out infrastructure priorities and identifies locational flexibility for hydrogen supply and demand options. Advances on limitations of previous research are made by developing an open-source model of the hydrogen system of Great Britain, based on three Net Zero scenarios set out by National Grid in their Future Energy Scenarios, in high temporal and spatial resolution. The model comprehensively covers demand sectors and supply options, in addition to extending the locational considerations of the Future Energy Scenarios. This study recommends prioritising the establishment of green hydrogen hubs, in the near-term, aligning with demands for synthetic fuels production, industry, and power, which can facilitate the subsequent roll out of  up to 10GW of hydrogen production capacity by 2050. The analysis quantifies a high proportion of hydrogen supply and demand which can be located flexibly.

\end{abstract}


\begin{highlights}
\item Analysis of hydrogen infrastructure under three future pathways for GB
\item Open-source model for transparent, reusable analysis of future GB hydrogen system
\item Recommend hydrogen hubs for synthetic fuels, industry, and power generation in near-term
\item High degree of locational flexibility of hydrogen supply and demand sites is quantified
\end{highlights}


\end{frontmatter}


\section{Introduction}
\label{}

Hydrogen is recognised globally as having a vital role in the decarbonisation of energy systems due to its ability to cover periods where renewable energy sources cannot meet demands or decarbonisation cannot be achieved with direct electrification \cite{[1]chapman2019review}. The 2022 global energy crisis increased the interest for low-carbon hydrogen economy development in multiple countries with the aim not only to reduce the emissions but also to decouple from fossil fuels dependency and diversify energy supply mix \cite{[2]IEA2022-GlobalH2}. Even though Liquefied Natural Gas (LNG) is currently the primary solution to meet short-term energy security in the European Union (EU) \cite{[62]pedersen2022long}, hydrogen facilities development can be seen in multiple countries around the world including export and import ports \cite{[61]chen2023review} with intend to diversity energy supply sources in a long-term run. 

The United Kingdom's (UK’s) abundant onshore and offshore wind resources \cite{[3]BEIS2022-Wind} could facilitate the realisation of the UK government's hydrogen ambitions, which includes 10GW of low-carbon hydrogen production by 2030 \cite{[4]BEIS2022-H2StrategyDecember}. This hydrogen could be used for decarbonisation of various UK energy sectors in addition to the potential to export to continental Europe \cite{[4]BEIS2022-H2StrategyDecember},\cite{[5]HMGov2023-H2Roadmap}. However, current lack of clarity in projections for hydrogen market development holds back investment decisions which in turn slows down the growth of low-carbon hydrogen supply chain.

The Future Energy Scenarios (FES) by National Grid \cite{[6]ESO2023-Report} are comprehensive whole energy system projections which include potential pathways for the integration and adoption of hydrogen. These scenarios are meticulously crafted through a combination of extensive research, data analysis, and industry insights to forecast plausible developments in hydrogen use within the energy system.

The basis of these scenarios lies in examining various factors, such as technological advancements, policy changes, market dynamics, and societal shifts, to outline possible future trajectories for hydrogen utilisation. This includes assessing the potential demand for low-carbon hydrogen across different sectors like transportation, industry, and residential use, along with the necessary infrastructure requirements. Therefore, FES provides a basis for potential development of low-carbon hydrogen economy from minimum to maximum adaptation, covering all potential sensitivity cases in between.

FES is suitable to be used to answer many questions around future hydrogen infrastructure development. However, the FES underlying modelling is not open-source and it is not easy to analyse the impacts of sensitivities to assumptions of hydrogen infrastructure in demand sectors, supply options, and pipelines, in addition to the spatial distribution of this infrastructure. Therefore, there is a need for a specific model of the hydrogen energy system which can use FES as the baseline with added functionality to analyse sensitivity of assumptions important for understanding the evolution of hydrogen infrastructure in the UK.

\subsection{Hydrogen in Future Energy Systems in the UK and EU}

Recent research on hydrogen infrastructure development in the UK has aimed to provide informed projections and has ranged from initial feasibility studies to the more detail national assessments. However, these studies are limited in their coverage of demand sectors and supply options, and are limited in spatial representation.

Modelling of the spatial expansion of hydrogen infrastructure in the UK within the context of broader energy system interactions was undertaken by \citet{[43]balta2013spatial}, in 2013. This study proposed a geographical infrastructure development plan with ten-year intervals, spanning from 2020 to 2050. However, this study did not cover all sectors, only	heating and transport (road and aviation) on the demand side, and mainly disaggregated locational aspects based on geographic population density. Additionally, scenarios were based on different levels of emission reduction by 2050 with high reliance on liquid hydrogen import.

Research conducted by \citet{[44]moreno2017towards}, in 2017, proposed hydrogen network development in stages starting from Midlands and expanding across the UK by 2070. The study incorporated the development of carbon dioxide (CO$_2$) pipeline infrastructure from methane reformation to offshore storage. This study did not cover all sectors, only hydrogen demand for road transport is included (in view of filling stations), while electrolysis is only included as source of hydrogen production in 2060.

\citet{[15]samsatli2019role}, in 2019, covered crucial locational elements such as electrolysers location and investigate trade-offs between energy transport via the transmission grid or hydrogen pipelines. This model accounted for inter-seasonal and daily demand fluctuations, as well as the intermittent nature of renewable power generation. Furthermore, it assessed various subsidy levels for upgrading heating devices within the system optimization framework. The study only covered limited demand sectors, the	demand side covered consumption by only the heating sector. There were also limited options for supply, e.g., hydrogen production from methane reformation, biomass gasification, and nuclear electrolysis were not covered.  

A recent study, by \citet{[45]akhurst2021tools} as part of the ELEGANCY project, focused on the transportation and storage of hydrogen and carbon dioxide from the methane reformation process, with the proposal to phase out investment in hydrogen infrastructure and adjust it based on results of the first phase outcome. However, the study covered limited demand sectors, the demand side only covers consumption by heating and transport. Supply options were limited; hydrogen produced mainly by methane reformation with carbon capture including small share of biomass gasification, and water electrolysis was suggested only to cover peak hydrogen demand in case hydrogen storage not available. 

All of these studies provide a useful basis for further research and developments. However, the developed models have not been released as open access and therefore new studies have to start modelling from ‘scratch’ investing substantial amount of time for data collection and validation. The ELEGANCY project \cite{[45]akhurst2021tools} reports to use an open-source software, but the model could not be located to perform a critical analysis of the assumptions made.

PyPSA-Eur is a EU-wide model including hydrogen infrastructure \cite{[46]neumann2023potential} which covers all sectors, with explicit locational data, and is open-source. PyPSA-Eur contains data for historical years and uses whole system least-cost optimisation to generate future scenarios. However, it is not trivial to incorporate data from pre-defined future scenarios, such as National Grid's FES which has been developed to align with individual UK government targets and specific UK regional developments. PyPSA-Eur contains valuable open-source data to model a hydrogen network anywhere in Europe, however, for this study the authors decided it was more suitable to build a new model than incorporate new data into PyPSA-Eur from National Grid's FES.

Overall, research using models to understand the development of hydrogen infrastructure development in the UK would benefit from methods with higher coverage of both demand sectors and supply options, in addition to improved spatial representation.

\subsection{Aim and Contributions}

This study aims to model and evaluate the development of Great Britain’s (GB’s) hydrogen infrastructure to (i) identify investment priorities, and (ii) assess locational flexibility. This study will analyse the minimum and maximum low-carbon hydrogen demand scenarios across the three net zero scenarios in FES in order to pinpoint key dependencies and establish resilient policies, ensuring informed and 'no-regret' investment decisions. It will also quantify the locational flexibility of hydrogen supply and demand using FES as the baseline, and discuss the risks associated with proceeding in different directions in the decarbonisation of the energy system.

In order to answer these aims, this paper will describe the developed model of the future GB hydrogen system incorporating all producers and consumers included in FES \cite{[8]ESO2022-Workbook}, in addition to locational aspects. Other essential inputs were sourced from PyPSA-GB \cite{[11]lyden4509311open} - a future power system model of GB based on FES. This was done with explicit locational aspects included. 

While policies are for the UK, the scope of this study is limited to GB only, representing England, Wales, and Scotland, and excludes Northern Ireland which has an integrated electricity market with the Republic of Ireland.

The novel contributions of this study are: 

(1) Investment priority near-term recommendations to support 'no-regret' investment decisions based on comparative analysis of FES decarbonisation scenarios and on enabling the pathway future infrastructure scale.

(2) Quantification of the locational flexibility of demand sectors and supply options, building on the locational analysis in FES.

(3) An open-source model, PyPSA-GB-H2, covering all potential producers and consumers of low-carbon hydrogen in future GB energy system as forecast in the FES. 

\section{Methodology}
\label{}

This section outlines the methodology utilised in this study, summarising details of the developed model, scenarios selection, assumptions made for hydrogen supply and demand modelling, and the network modelling approach, while also addressing the study's limitations with potential areas for future improvements.

\subsection{Open-source Model for GB Future Hydrogen System}

PyPSA-GB-H2 \cite{[63]PyPSA-GB-H2-2023} available at \href{https://github.com/tatyanadergunova/PyPSA-GB-H2.git}{https://github.com/tatyanadergunova/PyPSA-GB-H2.git}, includes data and models for hydrogen production, consumption, and storage. It was built using the open-source energy system framework PyPSA \cite{[58]brown2017pypsa}, which can model whole energy systems, including hydrogen components for production, storage, transport, and demand. It represents hydrogen flows using linear representations as energy streams, without considering more complex non-linear behavior.  This allows users to analyse different energy system scenarios by using linear programming to quickly optimize operation and investments to minimize costs.

A challenge in future energy systems modelling is dependence on non-linear network optimization, which often introduces numerous uncertainties and limitations, as emphasized by \cite{[47]mavromatidis2018review}. In addition, limited scenarios variety does not represent all potential pathways of energy systems development making review of the whole system narrowed down to boundaries set by multiple uncertainties.

Early development of hydrogen infrastructure needs to account for long-term goals and therefore lay the foundations for capacity targets out to the year 2050, in line with net zero goals. For the scenarios evaluated for year 2050, the sections of future infrastructure which would experience load in all potential scenarios development could be identified. These sections will be utilised in future regardless of selected scenario of hydrogen adaption and ‘early’ development of infrastructure around them could be considered as ‘no regret’ decision. Therefore, this study covered year 2050 only.

\subsection{FES Data and Modelled Scenarios}

National Grid's FES scenarios were used as the basis for data across the modelled hydrogen infrastructure. Hydrogen energy flow estimates from these reports are available to public and were used as the foundation for setting hydrogen production and demand capacities in PyPSA-GB-H2 model. Only FES-2022 workbook data \cite{[8]ESO2022-Workbook} was run in this study and further studies can be updated utilizing FES-2023 workbook as an input of low-carbon hydrogen consumption in each scenario \cite{[6]ESO2023-Report}. FES reports covering four different scenarios; however, the scope of this study is limited to evaluation of the three net zero scenarios.

This section discusses each of the hydrogen components modelled: supply, demand, network, and storage.

\subsubsection{Supply}

Approximately 2.5GW of carbon-intensive hydrogen production capacity is currently installed in the UK with about two-thirds coming as by-product from industrial processes (mainly steelworks and refineries) and only one-third being produced by methane reformation without carbon capture \cite{[12]BEIS2022-H2StrategyJuly}.

To support future development of low-carbon hydrogen supply options, the UK government's current approach supports two major types of hydrogen production: 1) electrolytic hydrogen production; and 2) methane reformation with carbon capture utilization and storage (CCUS) \cite{[4]BEIS2022-H2StrategyDecember}. 

At this stage of development, electrolysers have to be connected directly to renewable energy sources (`non-networked’ electrolysis) as the power grid is yet to be decarbonized to introduce `networked' electrolysis as highlighted by \cite{[23]mac2021hydrogen}. Another type of electrolytic hydrogen production reviewed by FES-2022 \cite{[8]ESO2022-Workbook} is ‘nuclear’ electrolysis, where more uncertainties could be seen due to ageing nuclear reactors \cite{[24]WorldNuclear2023},\cite{[25]novak1987nuclear} and criticism from public on new nuclear power development \cite{[26]spence2010public}. 

This section describes the hydrogen supply data in relation to the efficiency, availability, hourly production and spatial locations.

\subsubsection*{\textbf{Efficiency and Availability}}

Hydrogen production technologies have two major parameters for evaluation of actual output capacity - efficiency and availability. Efficiency is defined by the ratio of useful energy output to the total energy input. While availability, also known as the average capacity factor, is associated with requirements of planned equipment maintenance and feed stock supply pattern and can be defined as ratio of produced energy to production capacity. 

FES-2022 workbook \cite{[8]ESO2022-Workbook} was reviewed and efficiency and availability for each hydrogen production technology evaluated which is summarized in Table \ref{tab:efficiency}. 

\begin{table}[h]
\centering
\captionsetup{textfont=it, skip=0.5ex}
\caption{Efficiency and availability of hydrogen production technologies (FES-2022) \cite{[8]ESO2022-Workbook}}
\label{tab:efficiency}
\begin{tabular}{lcc}
\cline{1-2}
\rowcolor[HTML]{000000} 
{\color[HTML]{FFFFFF} \textbf{Technology}} &
  \multicolumn{1}{l}{\cellcolor[HTML]{000000}{\color[HTML]{FFFFFF} \textbf{Availability}}} &
  {\color[HTML]{FFFFFF} \textbf{Efficiency}} \\ \hline
\rowcolor[HTML]{EFEFEF} 
Methane Reformation with CCUS & 95\%    & 74-82\% \\
\rowcolor[HTML]{EFEFEF} 
Biomass Gasification          & 90\%    & 71-78\% \\
\rowcolor[HTML]{EFEFEF} 
`Nuclear' Electrolysis        & 80\%    & 84-88\% \\
\rowcolor[HTML]{EFEFEF} 
`Non-networked' Electrolysis  & 30\%    & 84-88\% \\
\rowcolor[HTML]{EFEFEF} 
`Networked' Electrolysis      & 44-50\% & 84-88\% \\ \hline
\end{tabular}
\end{table}

\emph{ Methane Reformation with CCUS \& Biomass Gasification.}	Efficiency of hydrogen production is constant throughout the years but varying in different scenarios, increasing with higher share of this hydrogen type in total mix. 

\emph{`Nuclear' Electrolysis, `Non-networked' Electrolysis \& `Networked' Electrolysis.}	Electrolysis efficiency is increasing through 2030-2050, assuming wider adaptation and technology maturity increase.

\emph{`Networked' Electrolysis.} Availability of ‘networked’ electrolysis is increasing through 2030-2050 assuming increase in flexibility and robustness of energy supply sources to the power grid. However, it stays steady at 40\% throughout all years in ‘System Transformation’ scenario with high reliance on hydrogen for heating and high share of methane reformation with CCUS. 

\subsubsection*{\textbf{Hourly Production}}

For this study hourly fluctuations of hydrogen production by methane reformation with CCUS, biomass gasification and ‘nuclear’ electrolysis will be neglected.  The major fluctuation in these methods of hydrogen production is associated with equipment maintenance requirements which are typically performed on schedule and could be planned in sequence. Therefore, steady-stage hourly hydrogen production throughout the year is deemed to be sufficient for country scale simulation disregarding efficiency and availability. 

For hourly production of ‘networked’ and ‘non-networked’ electrolysis, the actual data of renewable energy generators extracted from PyPSA-GB \cite{[11]lyden4509311open} will be adopted to cover hourly fluctuations. Annual average availability of offshore and onshore wind was reviewed, and it was recognized that it has similarities with ‘networked’ and ‘non-networked’ electrolysis, standing at about 50\% and 30\% accordingly. Based on this, offshore wind availability was assigned to ‘networked’ electrolysis and onshore wind to ‘non-networked’ electrolysis. Full integration of PyPSA-GB-H2 to PyPSA-GB can be completed in future research, while will result in more accurate availability of electrical power for ‘networked’ electrolysis.  

\subsubsection*{\textbf{Spatial data}}

To account for spatial considerations, each technology dependency on feedstock supply and other requirement were reviewed with summary presented in Table \ref{tab:geospatial}. 

\begin{table}[h]
\centering
\captionsetup{textfont=it, skip=0.5ex}
\caption{Geographical limitations of hydrogen production technologies by industrial clusters}
\label{tab:geospatial}
\begin{tabular}{lc}
\hline
\rowcolor[HTML]{000000} 
{\color[HTML]{FFFFFF} \textbf{Technology}} & {\color[HTML]{FFFFFF} \textbf{Limited to Cluster}}                                                                          \\ \hline
\rowcolor[HTML]{EFEFEF} 
Methane Reformation with CCUS              & \begin{tabular}[c]{@{}c@{}}St. Fergus, Grangemouth, Teesside, Humberside, Bacton, \\ Merseyside, Theddlethorpe\end{tabular} \\
\rowcolor[HTML]{EFEFEF} 
Biomass Gasification                       & \begin{tabular}[c]{@{}c@{}}Bacton, Grain LNG, Southampton, \\ South Wales, Theddlethorpe\end{tabular}                       \\
\rowcolor[HTML]{EFEFEF} 
`Nuclear' Electrolysis       & Bacton, Grain LNG, South Wales, Merseyside \\
\rowcolor[HTML]{EFEFEF} 
`Non-networked' Electrolysis & St. Fergus, Grangemouth, Merseyside        \\
\rowcolor[HTML]{EFEFEF} 
`Networked Electolysis       & Proposed location in FES-2022 \cite{[8]ESO2022-Workbook}          \\
\rowcolor[HTML]{EFEFEF} 
Imports                      & Bacton, Grangemouth                        \\ \hline
\end{tabular}
\end{table}

\emph{Methane Reformation with CCUS.} Offshore CO$_2$ storage is assumed at similar locations as hydrogen storage potential sites investigated by \cite{[32]peecock2023mapping}. Therefore, proximity to depleted oil and gas fields is assumed as beneficial location for steam reformation with CCUS to reduce CO$_2$ transport infrastructure. 

\emph{Biomass Gasification.}	Limited to the south part of England and Wales, areas of higher population density as no major pre-requisites were identified at this stage. 

\emph{`Nuclear' Electrolysis.}	Civilian nuclear fleet of the UK is ageing, and currently operated nuclear reactors are projected to be decommissioned in coming years \cite{[24]WorldNuclear2023},\cite{[25]novak1987nuclear}. Therefore, for the purpose of this study potential location of proposed nuclear power plants was assumed as location of future ‘nuclear’ electrolysis. 

\emph{`Non-networked' Electrolysis.} Potential locations of ‘non-networked’ electrolysis were assumed in Scotland and North-West of England.

\emph{`Networked' Electrolysis.} The given values in FES-2022 workbook \cite{[8]ESO2022-Workbook} represent the final installed capacity of ‘networked’ electrolysis by region and include ‘nuclear’ electrolysis. It is assumed that all indicated capacity is ‘networked’ electrolysis and generation is split between industrial cluster based on share for each region. Optimization of ‘networked’ electrolysis location shall be done in further studies after PyPSA-GB and PyPSA-GB-H2 interconnection.

\emph{Imports.}	Based on Project Union export pipeline interconnections \cite{[9]ProjectUnion2022}. The location of potential import points through seaports is excluded from this study.

\subsubsection{Demand}

Hydrogen demand varies across sectors in the different scenarios. \emph{`Consumer Transformation'} scenario has the lowest hydrogen adaptation which is mainly driven by electrified heating. \emph{'Leading the Way'} scenario adopts mixed approach for residential heating with about 43TWh of hydrogen demand by this sector in 2050. Further, hydrogen consumption by industries increased from about 11TWh in `Consumer Tranformation' to about 54TWh for year 2050. \emph{`System Transformation'} scenario includes about 145TWh of hydrogen for residential heating in 2050. In addition, industrial consumption increased to about 88TWh in 2050. The total hydrogen demand in 2050 for the three net zero scenaros is shown in Table \ref{tab:total demand}. 

\begin{table}[h]
\centering
\captionsetup{textfont=it, skip=0.5ex}
\caption{Total hydrogen demand in 2050 (FES-2022) \cite{[8]ESO2022-Workbook}}
\label{tab:total demand}
\begin{tabular}{lc}
\hline
\rowcolor[HTML]{000000} 
{\color[HTML]{FFFFFF} \textbf{Scenario}} & \multicolumn{1}{l}{\cellcolor[HTML]{000000}{\color[HTML]{FFFFFF} \textbf{Annual Hydrogen Demand in 2050, TWh}}} \\ \hline
\rowcolor[HTML]{EFEFEF} 
Consumer Transformation & 113 \\
\rowcolor[HTML]{EFEFEF} 
Leading the Way         & 244 \\
\rowcolor[HTML]{EFEFEF} 
System Transformation   & 431 \\ \hline
\end{tabular}
\end{table}

The following demand sectors are reviewed as potential low-carbon hydrogen consumers: heating, industry, power generation, transport, blending, and exports. Ranges of annual demand for each demand sector are from FES \cite{[8]ESO2022-Workbook}.

\subsubsection*{\textbf{Heating}: 0-145TWh H$_2$ in 2050}

Heating decarbonization has one of the most debatable paths, with extensive competition between direct electrification (heat pumps) and ‘clean’ gas (hydrogen boilers). From one hand, the process efficiency of hydrogen boilers in heating sector is significantly lower than that of heat pumps, accounting for about 65\% and 90\% respectively \cite{[13]ESO2022-Report}. This fact creates an open debate for delay in policymaking decision and indirect support of less attractive techno-economic solution as hydrogen according to \cite{[14]lowes2020heating}. However, \cite{[15]samsatli2019role} highlights potential constraints of multi-family house (apartment blocks) heating decarbonization by ground-source heat pumps and economic downsides of air-source heat pumps which could potentially be used in this application. Another option for apartment blocks heating is district heating which could be fed by ground-source heat pumps as summarized by \cite{[60]david2017heat}, however, techno-economic feasibility of this option is still to be confirmed.

Heating creates substantial seasonal fluctuations and its crucial to model it dynamically covering not only daily fluctuations but also hourly what will allow evaluation of peak storage requirement and infrastructure sizing. Natural gas is still the primary source for heating in the UK, accounting for 74\% of households in 2021 according to \cite{[50]UKParl2023-Heating}. Therefore, share of natural gas consumed by each Local Distribution Zone (LDZ) \cite{[51]NationalGas2023-Transmission} was split between industrial clusters obtaining share of total consumption per each cluster. These shares were further used to calculate annual demand of each cluster based on total annual hydrogen consumption for heating (residential and commercial) per FES-2022 workbook \cite{[8]ESO2022-Workbook}. For hourly heat demand fluctuations hourly ambient temperature fluctuations and type of buildings (single family house, multifamily house or commercial) were used as an input to produce annual heat demand per each type of building, see \cite{[52]PyPSA2023} for more details. 2010 temperature profile of the UK was assumed for this study. The proportion of single (~80\%) and multi-family (~20\%) houses is adopted for 2021 from \cite{[53]OfficeNatStat2023} and provided as annual heat demand input per buildings type. As output data will be used to obtain share only, exact heat demand is not considered as critical and assumed similar to given in \cite{[52]PyPSA2023}. Annual temperature readings were taken from Renewables Ninja \cite{[54]RenewablesNinja2023}. 

\subsubsection*{\textbf{Industry}: 11-88TWh H$_2$  in 2050}

The UK industries are concentrated in ‘clusters’ and its decarbonization strategy was issued in March 2021 by BEIS (Department for Business, Energy \& Industrial Strategy) \cite{[16]HMGov2023-IndustrialDecarb}. Under this strategy a review of potential decarbonization pathways for each industrial sector is presented. The review shows that even though hydrogen can be applied in decarbonization of each industry to a certain extent, the highest potential of hydrogen application is in clustered sites.

To estimate energy consumption by industries, daily fluctuations of natural gas consumption by industries from National Gas transmission data \cite{[51]NationalGas2023-Transmission} were used. The obtained share was then divided by 24 hours and multiplied by total annual energy demand by industry sector from FES-2022 workbook \cite{[8]ESO2022-Workbook}.  Further, all consumption were assigned per industrial clusters adopted for this study. Even though the roadmap for industrial decarbonization is not defined yet and may account for some share of electrification, the selected approach is deemed to be acceptable for this level of study.

\subsubsection*{\textbf{Power Generation}: 12-17TWh H$_2$ in 2050 \cite{[8]ESO2022-Workbook}}

To cover variations in renewable electricity generation, a long-term energy storage is critical. Different technologies were proposed for energy storage to cover peak electricity demand and curtailment reduction; however, a single type of storage cannot provide sufficient system flexibility as indicated by \cite{[17]cardenas2021energy}. According to evaluation by \cite{[18]AFRY-for-CCC2023} performed for Climate Change Committee (CCC), power generation sector will be the major hydrogen consumer in the nearest future with the UK government commitment of electricity grid decarbonization by 2035 (subject to energy security). 

For hydrogen power generation hourly output from PyPSA-GB \cite{[11]lyden4509311open} was used and power plant’s location split between clusters based on geographical proximity. 

\subsubsection*{\textbf{Transport}: 83-138TWh H$_2$ in 2050 \cite{[8]ESO2022-Workbook}}

Transport sector covers road transport, shipping, aviation, and rail. Hydrogen as fuel is not projected to receive widespread adaptation for light duty vehicles as per \cite{[19]ESO2022-CallforEvid}, while decarbonization of shipping and aviation is projected to utilise hydrogen equally in all FES decarbonization scenarios in 2050 accounting for about 70TWh and 10TWh respectively \cite{[8]ESO2022-Workbook}. According to \cite{[20]gray2021decarbonising} hydrogen will be used as feedstock for synthetic fuels production for shipping and aviation industries rather than a fuels in pure form. As for rail sector, is more likely to proceed with electrification and maximum annual consumption is predicted to be limited to 2TWh in 2050 \cite{[8]ESO2022-Workbook}.

Share of road transport energy consumption by regions was taken from \cite{[55]GovUK2022-NatStatist} and multiplied by total energy consumption from FES-2022 workbook \cite{[8]ESO2022-Workbook}. Hourly hydrogen consumption assumed to be steady throughout the year considering buffer storage of liquid hydrogen at consumer side.  

For shipping and aviation, hydrogen consumption split among clusters based on location of airports and major seaports is not accurate as future fuel for this transportation may not be hydrogen itself but rather its derivatives in form of synthetic fuels according to \cite{[20]gray2021decarbonising}. Therefore, hydrogen consumption shall be related to synthetic fuel production areas which presumably could be in existing industrial clusters. Based on this, shipping and aviation demand was combined.  Hourly hydrogen consumption assumed to be steady throughout the year as chemical plants have controlled production and synthetic fuels buffer storage can be provided for end users demand fluctuation. The same assumption was adopted for rail as it has relatively low demand compared to other consumers.

\subsubsection*{\textbf{Blending}}

One of the short-term solutions for expansions of hydrogen market analysed by the UK government is hydrogen blending into natural gas networks with up to 20\% (by volume), with plans to announce the final decision on this approach in 2023 \cite{[4]BEIS2022-H2StrategyDecember}. Implementation of this strategy may create flexible consumption that is favourable condition to support low-carbon hydrogen production development according to \cite{[21]quarton2020should}. 

Even though the hydrogen blending into natural gas network is sill pending the UK government decision, FES-2022 indicate its phase-out by 2043 at the latest, meaning natural gas network will not be acting as hydrogen sink anymore. Therefore, this demand type was excluded from this study. This may be added in future studies where years earlier than 2050 of hydrogen deployment will be simulated.  

\subsubsection*{\textbf{Exports}}

At this early stage, the UK exploring opportunities to export hydrogen to Europe given its long relationships with energy trade and increasing hydrogen demand \cite{[4]BEIS2022-H2StrategyDecember}. Two potential export interconnectors were identified: 1) Bacton industrial cluster to Netherlands; 2) South-West border of Scotland to Ireland \cite{[6]ESO2023-Report}. However, the development of low-carbon hydrogen export projects will depend not only on production capability of the UK but also readiness of the import countries and economic viability.

FES-2022 does not include exports in demand side \cite{[8]ESO2022-Workbook}, therefore supply-demand balance would not be achieved if exports included in PyPSA-GB-H2 model. Further, FES-2023 \cite{[6]ESO2023-Report} excluded hydrogen exports from all scenarios in workbook, and there is only a mention of potential hydrogen exports. Based on above, exports were excluded from PyPSA-GB-H2 model and this shall be addressed in further studies. 

\subsubsection*{\textbf{Direct Air Carbon Capture and Storage}}

Hydrogen consumption by Direct Air Carbon Capture and Storage (DACCS) is present only in ‘Leading the Way’ scenario of FES-2022 standing at about 25TWh in 2050 \cite{[8]ESO2022-Workbook}. The purpose of DACCS described in FES reports \cite{[6]ESO2023-Report},\cite{[13]ESO2022-Report} is to create a negative emissions source by capturing CO$_2$ from air and storing it at local storage sites. It is also suggested in FES reports that DACCS shall be located close to industrial carbon capture and storage infrastructure, therefore similar approach to location as of methane reformation with CCUS is adopted. Even though there are different ways to obtain heat for DACCS, one of which is hydrogen, this study is not performing research on DACCS technological advancements and only evaluated hydrogen consumption by National Grid was used as a sink of low-carbon hydrogen to sustain DACCS energy needs.

\subsubsection{Network}

The existing gas infrastructure of the UK is not suitable for 100\% hydrogen use and substantial investments will be required in case of decision in its favour. Project Union \cite{[9]ProjectUnion2022} is performing a study in which existing pipelines infrastructure repurpose will be covered along with proposal of new pipelines construction. One of the limitations for the existing infrastructure repurpose is transition period where natural gas will still play a significant role and rely on its historic infrastructure. 
Alongside with Project Union, feasibility study for low-carbon hydrogen infrastructure development on a regional level is also performed being: 1) North-East Network and Industrial Cluster Development (Operator: SGN) \cite{[27]SGN2021}; 2) East-Coast Hydrogen (Operator: NGN, Cadent) \cite{[28]NGN2021}; 3) Capital Hydrogen (Operator: SGN, Cadent) \cite{[29]Cadent2022-CapitalH2}; 4) HyNet North West Hydrogen Pipeline (Operator: Cadent) \cite{[30]Cadent2022-HyNet}. Based on the above ongoing studies review, it is concluded that early development of hydrogen economy and phasing selection by specific region will depend not only on resource availability (renewable energy or natural gas) but also convenience for hydrogen storage and suitability of infrastructure for CO$_2$ storage in case of methane reformation with CCUS.  

Hydrogen pipelines between industrial clusters were modelled as energy flows, disregarding thermodynamic properties and loss of energy associated with frictional pressure drop. This simplification limits possibility of hydrogen infrastructure costs evaluation which shall include not only pipelines installation, but also compression stations and operational costs of electricity demand required for compression which can vary between 0.81-1.1 kWh/kgH$_2$ according to \cite{[48]niermann2021liquid}. Furthermore, estimated operating pressure (average) of salt caverns by \cite{[33]williams2022does} is 132-278 barg, what will need additional compression from hydrogen network operating at 40-80 barg \cite{[27]SGN2021},\cite{[28]NGN2021} and therefore capital and operational costs would increase further. 

\begin{figure}
\center
\footnotesize\emph{Modelled Hydrogen Network (Industrial Cluster Level)}
  \begin{center}
   \includegraphics[width=0.48\textwidth]{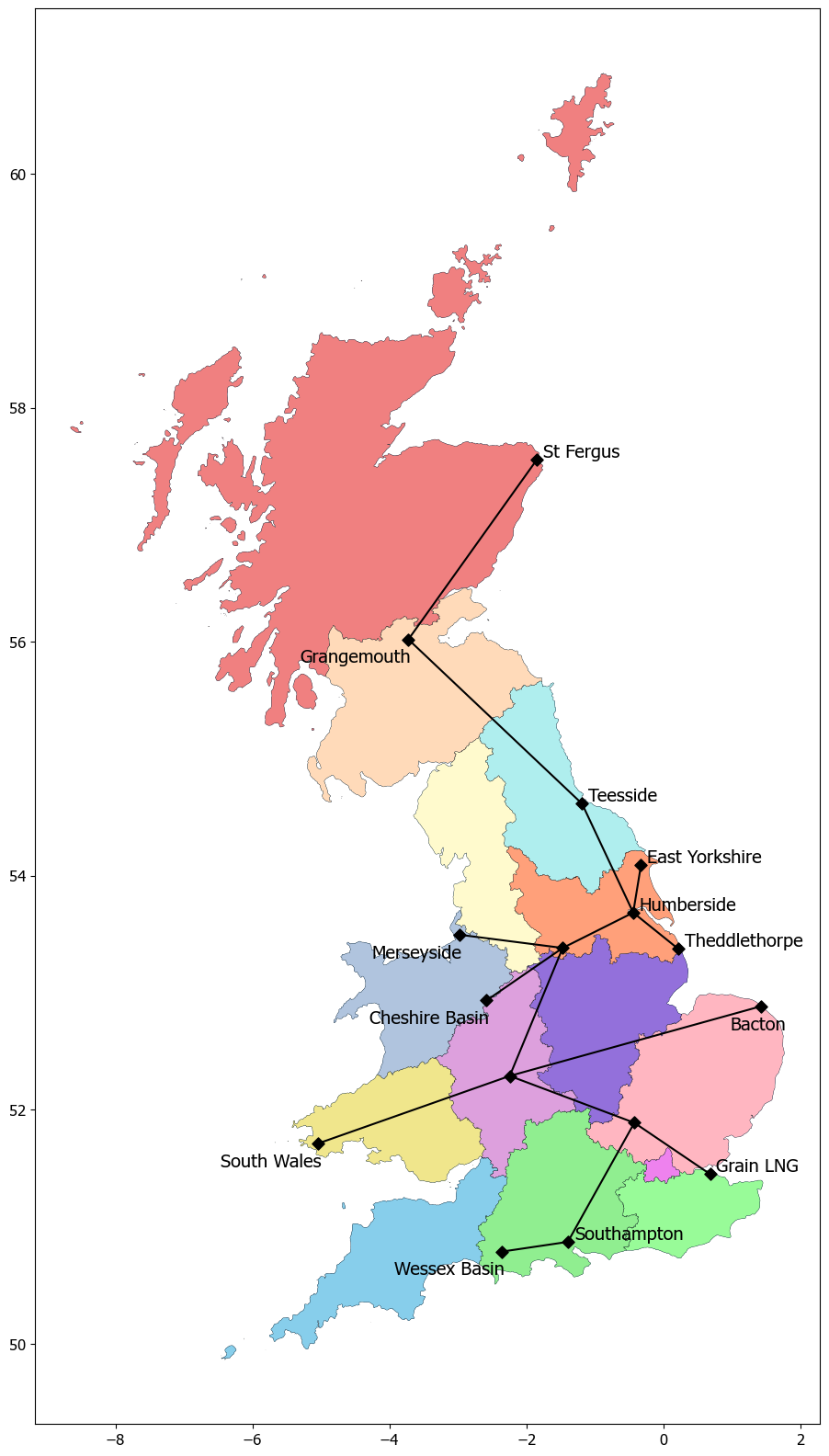}
  \end{center}
  \caption{\emph{Industrial clusters and salt caverns plot on GB Distributed Network Operator (DNO) License Areas map \cite{[64]ESO2023-DNO}. This Figure is for illustrative purposes only.}}
  \label{figure: network}
\end{figure}

Multiple models including \cite{[46]neumann2023potential}, model hydrogen transport as star network. However, according to hydrogen modelling evaluation performed by \cite{[49]reuss2019modeling} star network are not be representative for actual pipelines loads while tree network will benefit the study at low computational costs. Therefore, tree network model was adopted for this study, as shown on Figure \ref{figure: network}. The network simulation step of one hour was adopted to cover hourly fluctuations of supply and demand sides. 

This study introduced  geospatial division of hydrogen supply and demand and identified drivers for flexibility of sectors location on industrial clusters level. The assumption of industrial cluster level, shown on Figure \ref{figure: network}, serves the purpose of this study to identify the 'flexible' sectors to review minimum infrastructure requirement. However, higher resolution may be required for more detailed analysis. 

\subsubsection{Storage}

Another obstacle on the way of large-scale hydrogen deployment is storage, which will be vital to manage the difference in supply and demand patterns. Evaluation of theoretical capacity of offshore depleted hydrocarbon fields was done by \citet{[31]mouli2021mapping} and \citet{[32]peecock2023mapping} with potential storage capacity estimated at not less than 2,500TWh considering solely gas reservoirs. Further, study by \citet{[33]williams2022does} concluded that the UK salt caverns can accommodate about 2,000TWh theoretical hydrogen storage capacity. With maximum hydrogen storage requirement of about 55TWh in 2050 according to ‘System Transformation’ scenario peak \cite{[8]ESO2022-Workbook}, the estimated theoretical capacity of geological storage by far exceeding the requirements. Even though, the relatively high theoretical storage capacity compared to requirements could be taken as a positive sign, detailed research on hydrogen storage in porous media considering biological and geochemical reactions is still yet to be completed which may cause not only hydrogen contamination but also losses as emphasized by \citet{[34]heinemann2021enabling}.

Only salt caverns storage was assumed for this study considering potential contamination and competition with CO$_2$ storage for offshore depleted oil and gas fields. As highlighted by \cite{[34]heinemann2021enabling} there is a risk of hydrogen contamination and biochemical reactions in porous geological media. Further, \cite{[56]hemme2018hydrogeochemical} identified that CO$_2$ presence in depleted gas fields will cause higher level of hydrogen conversion to CH4 via methanogenesis. In addition, report issued for CCC on power system decarbonization \cite{[18]AFRY-for-CCC2023} assumed only salt caverns and tanks as hydrogen storage technology. It shall be noted that peak of required hydrogen storage capacity per FES-2022 workbook \cite{[8]ESO2022-Workbook} is about 55TWh which is by far smaller than available theoretical storage capacity of the UK salt caverns estimated by \cite{[33]williams2022does} which is summarized in Table \ref{tab:storage}. 

\begin{table}[h]
\centering
\captionsetup{textfont=it, skip=0.5ex}
\caption{Theoretical storage capacity of the UK onshore salt caverns \cite{[33]williams2022does}}
\label{tab:storage}
\begin{tabular}{lc}
\hline
\rowcolor[HTML]{000000} 
{\color[HTML]{FFFFFF} \textbf{Region}} & {\color[HTML]{FFFFFF} \textbf{Combined Theoretical Storage of all caverns, TWh}} \\ \hline
\rowcolor[HTML]{EFEFEF} 
Cheshire Basin & 129  \\
\rowcolor[HTML]{EFEFEF} 
East Yorkshire & 1465 \\
\rowcolor[HTML]{EFEFEF} 
Wessex Basin   & 557  \\ \hline
\end{tabular}
\end{table}

\subsubsection{FES Limitations}

The major limitation of FES-2022 scenarios \cite{[8]ESO2022-Workbook} in judgement for hydrogen infrastructure scale is share of hydrogen supply and demand by regions and technologies, with 2050 ‘networked’ electrolysis data being the only exception. The 'networked' electrolysis is defined as hydrogen production via water electrolysis with electricity sourced from power grid; therefore, the location of this producer shall be assigned to a particular region only when techno-economics of other less flexible hydrogen production technologies, hydrogen infrastructure and power grid expansion are evaluated.

\section{Results}
\label{}

This section describes PyPSA-GB-H2 modelling results for each scenario and their comparison with FES-2022 trends. Further, analysis of hydrogen supply-demand flexibility based on spatial constraints is outlined for each scenario. 

\subsection{Hourly Supply-Demand Trends}

\begin{figure}[h!]
\footnotesize\emph{(A) - 'Consumer Transformation \space\space\space\space\space\space\space\space\space (B) - 'Leading the Way' \space\space\space\space\space\space\space\space\space (C) - 'System Transformation'}
\includegraphics[scale=0.395]{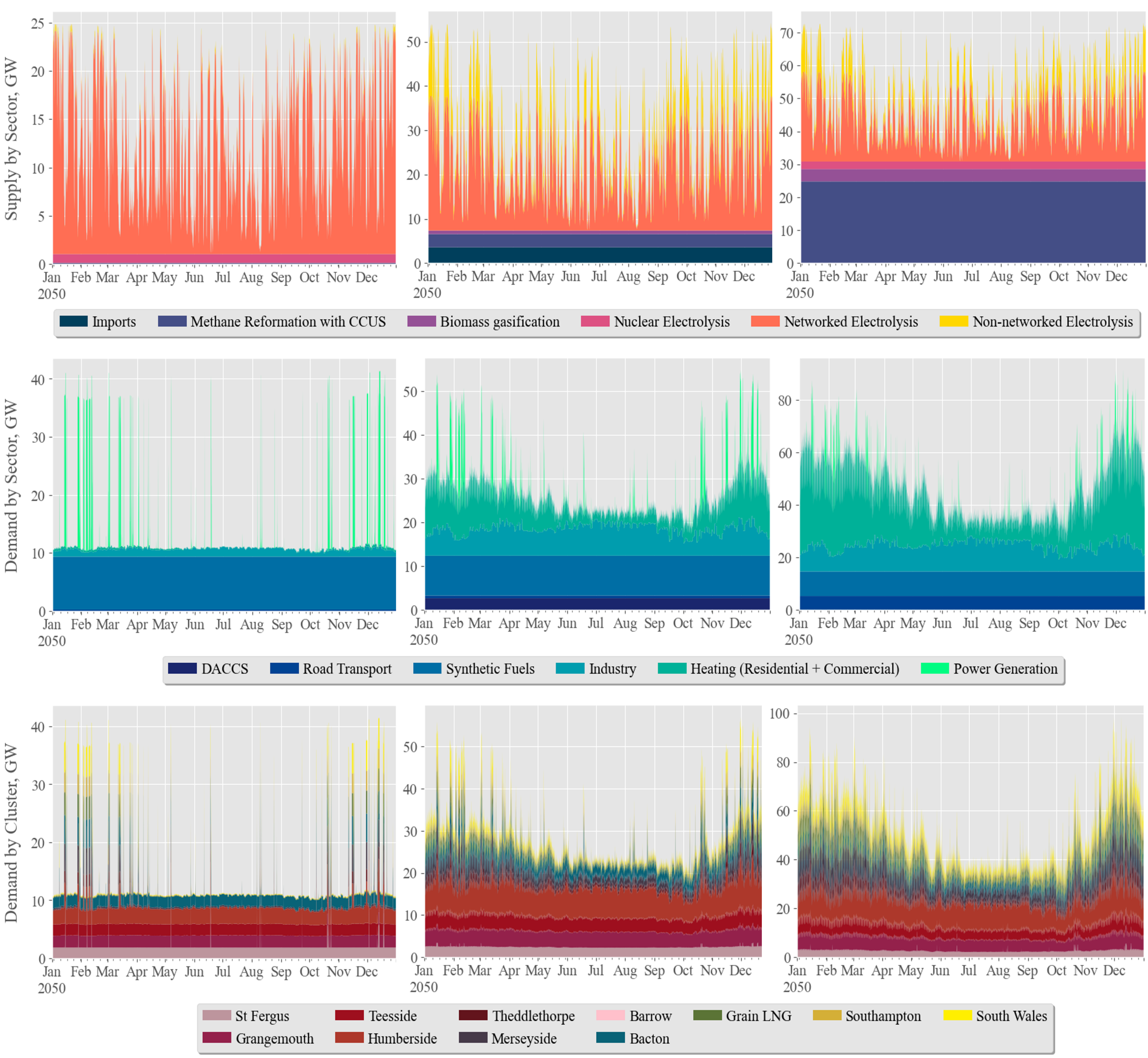}
\centering
\caption{\emph{Hourly hydrogen supply and demand by various sectors for three FES-2022 Net Zero scenarios, year 2050, based on PyPSA-GB-H2 output. Supply-demand patterns and cluster's share per methodology of this study.}}
\label{figure1}
\end{figure}

Output from PyPSA-GB-H2 for hourly supply-demand trends is presented in Figure \ref{figure1} for each FES-2022 decarbonization scenario.

\emph{'Consumer Transformation.'} Hydrogen supply in this scenario is predominated by ‘networked’ electrolysis at about 90\% of total production in 2050 followed by ‘nuclear’ electrolysis at about 7\% \cite{[8]ESO2022-Workbook}. Hydrogen production profile by ‘networked’ electrolysis resonates with offshore wind profile, while ‘nuclear’ electrolysis has a steady production. As for demand, it can be observed that the consumption throughout the year is relatively steady, except of high peaks in cold season (November-to-March). These sharp peaks are caused by hydrogen demand for power generation required to cover seasonal gaps in renewable energy production. Synthetic fuels (shipping and aviation) is the main consumer in this scenario standing at about 70\% in 2050 \cite{[8]ESO2022-Workbook}. 

\emph{'Leading the Way.'} The largest share of hydrogen supply in this scenario is ‘networked’ electrolysis accounting for about 50\%, followed by ‘non-networked’ electrolysis at about 20\% of total annual production in 2050 \cite{[8]ESO2022-Workbook}. ‘Networked’ electrolysis availability was adopted from offshore wind and ‘non-networked’ electrolysis from onshore wind profiles. Fluctuations of hydrogen consumption throughout the year are more pronounced in this scenario  compared to ‘Consumer Transformation’ as can be seen from Figure \ref{figure1} . This is caused by increased demand of hydrogen by heating (residential and commercial) and industrial sectors. Sharp consumption peaks by power generation are also observed in cold season. Hydrogen demand in this scenario predominated by three sectors: synthetic fuels production (shipping and aviation) about 30\% of total annual consumption in 2050, followed by heating (residential and commercial) and industry standing at about 25\% and 20\% respectively \cite{[8]ESO2022-Workbook}.

\emph{'System Transformation.'} Unlike other FES scenarios, hydrogen supply in ‘System Transformation’ largely come from methane reformation with CCU with about 50\% of total supply in 2050; ‘networked’ electrolysis still plays an important role with approximately 30\% of total supply \cite{[8]ESO2022-Workbook}. In addition, ‘System Transformation’ scenario has the highest fluctuations of demand throughout the year in comparison to other FES scenarios. This is caused by higher share of hydrogen for heating (residential) with about 30\% of total annual demand in 2050. Hydrogen consumption for synthetic fuels production (shipping and aviation) stays the same throughout all scenarios, with hydrogen to power generation slightly decreasing with total hydrogen consumption increase, therefore, high peaks becoming less dominant. 

\subsection{Clusters' Balance and Network Load}

\begin{figure}[h!]
\footnotesize\emph{(A) - 'Consumer Transformation \space\space\space\space\space\space\space\space\space (B) - 'Leading the Way' \space\space\space\space\space\space\space\space\space (C) - 'System Transformation'}
\includegraphics[scale=0.39]{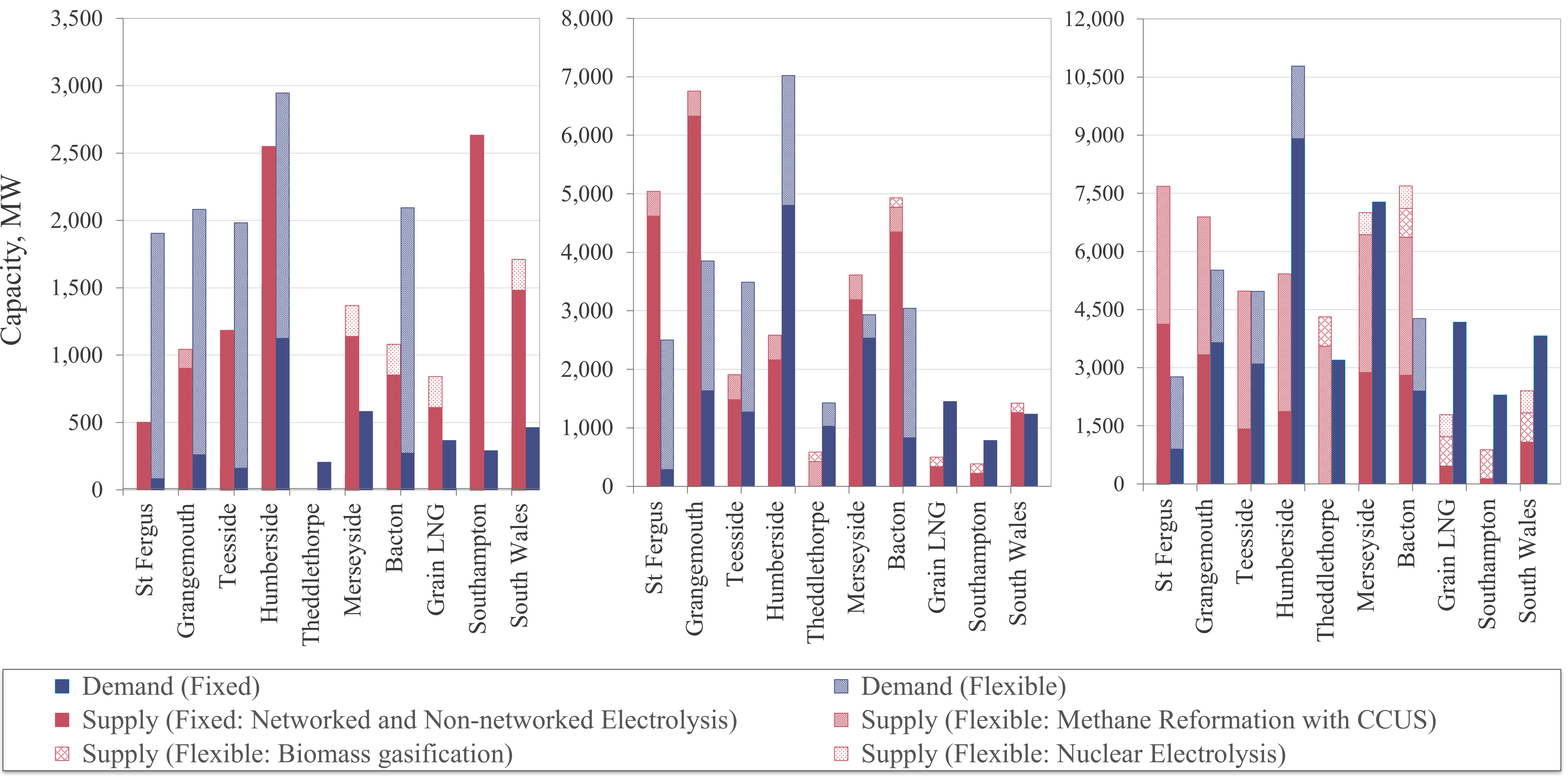}
\centering
\caption{\emph{FES-2022 scenarios supply-demand geospatial flexibility, year 2050. With increased share of hydrogen adaptation in GB’s energy system demand side flexibility is dropping bringing the requirement of complex hydrogen infrastructure. This is mainly caused by dispersed demand of residential heating and a high developed infrastructure will be required regardless the increase in supply flexibility.‘Networked’ electrolysis was assigned to 'fixed' locations as per FES-2022 workbook \cite{[8]ESO2022-Workbook}; however, further evaluation shall be done upon PyPSA-GB-H2 and PyPSA-GB \cite{[11]lyden4509311open} interconnection.}}
\label{figure2}
\end{figure}

\begin{figure}[h!]
\footnotesize\emph{(A) - 'Consumer Transformation \space\space\space\space\space\space\space\space\space (B) - 'Leading the Way' \space\space\space\space\space\space\space\space\space (C) - 'System Transformation'}
\includegraphics[scale=0.385]{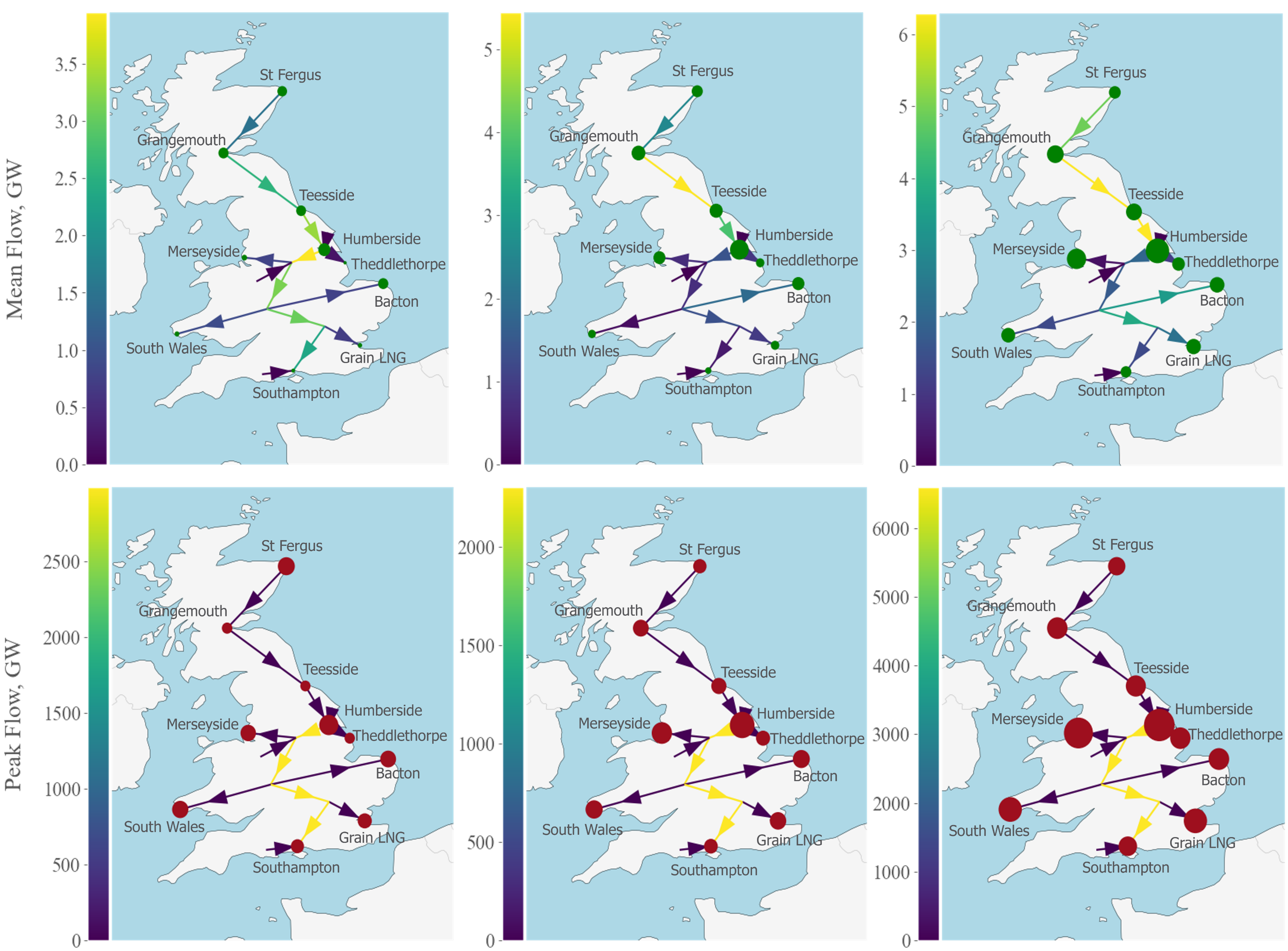}
\centering
\caption{\emph{PyPSA-GB-H2 output for mean and peak hydrogen energy flows including demand per cluster for three FES-2022 Net Zero scenarios, year 2050.}}
\label{figure3}
\end{figure}

Figure \ref{figure2} gives graphical representation of supply-demand spatial flexibility and Figure \ref{figure3} shows mean hydrogen flows between clusters for each FES scenario in 2050. 

\emph{‘Consumer Transformation.’} Even though this scenario has the lowest annual hydrogen consumption compared to other FES scenarios, hydrogen generation is not equal to consumption for each cluster. This is mainly related to modelling limitations in regards of spatial resolution and fixation of future ‘flexible’ demand such as synthetic fuels production. Demand flexibility in this scenario is governed by synthetic fuels production for shipping and aviation industries, while methane reformation with CCUS and ‘nuclear’ electrolysis represent supply flexibility and limited to certain clusters based on feedstock / additional infrastructure dependency. As for lines loading, a substantial difference between mean and peak hydrogen flows can be observed in pipelines connecting clusters and Battery Limits (B/Ls), the points where hydrogen flow splits to two or more clusters as can be seen on Figure \ref{figure3} . This is caused by high instantaneous flows of hydrogen to power generation and model resolution to industrial clusters rather than smaller regions like Grid Supply Points (GSPs).

\emph{'Leading the Way'} has the medium adaptation of hydrogen, compared to other FES-2022 scenarios. Figure \ref{figure2} shows the increased difference in supply-demand balance within one node causing higher hydrogen flows between the clusters. The higher share of ‘fixed’ supply and demand making ‘Leading the Way’ scenario less flexible compared to ‘Consumer Transformation’. Nonetheless, strategic location of synthetic fuels production (shipping and aviation) and DACCS may allow hydrogen infrastructure reduction to regional levels. The lines loading of ‘Leading the Way’ scenario have similar trends as of ‘Consumer Transformation’ having extreme loading caused by instantaneous flows to power generators what can be seen in Figure \ref{figure3}.

\emph{'System Transformation'} has the highest hydrogen adaptation particularly in heating sector, accounting for about 45\% of total hydrogen consumption by commercial and residential sectors in 2050 \cite{[8]ESO2022-Workbook}. This makes demand side less flexible in view of geographical distribution. Overall, there is still a space for infrastructure optimization with flexible consumers as synthetic fuels (for shipping and aviation) production and hydrogen producers as methane reformation with CCUS, biomass gasification and 'nuclear' electrolysis. 
‘System Transformation’ has the highest difference in mean and peak lines loading caused not only by power generation peaking but also hydrogen flow for residential heating due to low temperature peaks.

\subsection{Storage}

Overall, hydrogen storage state of charge follow similar trends for all scenarios in FES and PyPSA-GB-H2 model. However, there is a significant difference in the peak storage capacity. It is assumed that the observed difference is associated with modelling approach difference. The difference between required storage capacities can reduce or increase the total economic feasibility of low-carbon hydrogen adaptation. Therefore, further investigation is required as FES reports do not provide sufficient details of modelling.

\begin{figure}[h!]
\footnotesize\emph{(A) - 'Consumer Transformation \space\space\space\space\space\space\space\space\space\space (B) - 'Leading the Way' \space\space\space\space\space\space\space\space\space\space (C) - 'System Transformation'}
\includegraphics[scale=0.325]{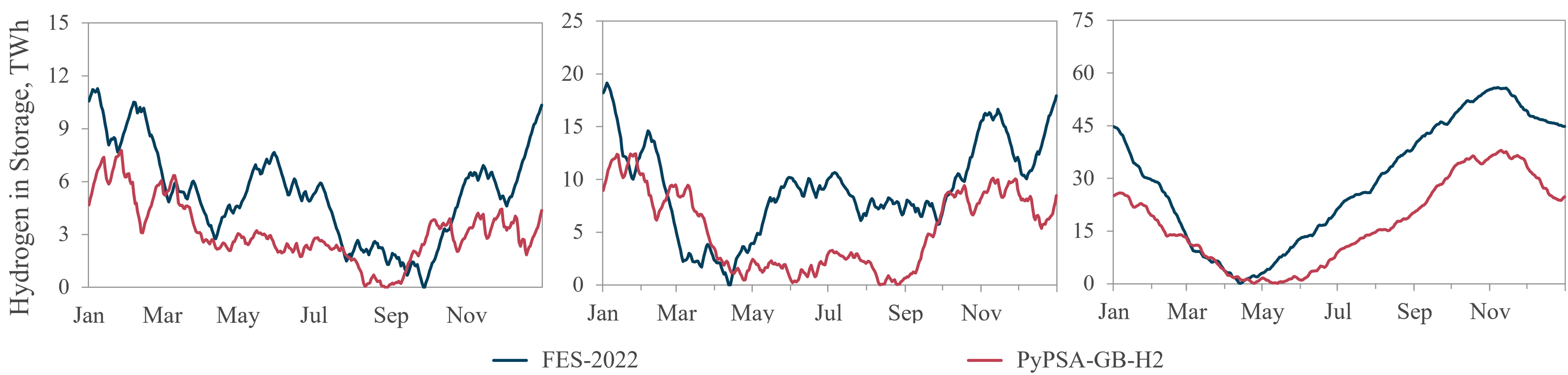}
\centering
\caption{\emph{Hydrogen storage 'state of charge' - FES-2022 \cite{[8]ESO2022-Workbook} vs PyPSA-GB-H2 \cite{[63]PyPSA-GB-H2-2023}, 2050. Even though a similar trends can be observed, both initial and peak storage capacity is lower in PyPSA-GB-H2 model compared to FES-2022, which is caused by difference in supply-demand modelling.}}
\label{figure:storage}
\end{figure}

\emph{'Consumer Transformation'.} The variation of hydrogen storage profile between FES-2022 and PyPSA-GB-H2 shown in Figure \ref{figure:storage} can be explained by the difference in supply-demand fluctuations. The major differences of PyPSA-GB-H2 from FES-2022 are high peaks of hydrogen demand in cold season (November-to-March) caused by hydrogen flow to power generation as modelling of this sector is based on power demand and does not account technological specifics of this process. In addition, hydrogen production curve in PyPSA-GB-H2 model has sharper peaks throughout the year associated with ‘networked’ electrolysis being major hydrogen producer and linked to offshore wind availability only.

\emph{'Leading the Way'.} Even though a similar trend can be observed, both initial and peak stage of charge is almost twice lower in PyPSA-GB-H2 model compared to FES-2022 (Figure \ref{figure:storage}) which is relevant to difference in supply-demand profile. Hydrogen demand by synthetic fuels (shipping and aviation) and power generation stays approximately at the same levels as of ‘Consumer Transformation’ scenario. The consumption spikes in FES-2022 could be associated with different approach to heating demand modelling: daily temperature profile and combined electric-hydrogen heating with hydrogen used for lower temperature peaks only. 

\emph{'System Transformation'.} Annual supply-demand profile for ‘System Transformation’ scenario in PyPSA-GB-H2 modelling is relatively close to FES-2022. FES-2022 demand curve is smoother compared to PyPSA-GB-H2 what could be caused by difference in temperature fluctuations andd introduction of local storage. Hydrogen storage state of charge is also showing high similarity between FES-2022 and PyPSA-GB-H2 models, with only variation in peak storage during cold season. 

\section{Discussions}
\label{}

As part of this study, an open-source model, PyPSA-GB-H2 \cite{[63]PyPSA-GB-H2-2023}, was developed to evaluate the requirement and scale of low-carbon hydrogen infrastructure to meet Great Britain's (GB's) Net Zero ambitions. The model incorporated all potential consumers and production technologies of low-carbon hydrogen proposed in Future Energy Scenarios (FES) \cite{[6]ESO2023-Report} which are being issued by National Grid based on collaboration with various stakeholders. The demand and production capacities in this study were derived from FES-2022 workbook \cite{[8]ESO2022-Workbook}, however, PyPSA-GB-H2 model can be used to run FES workbooks for other years, including FES-2023. Built upon low-carbon hydrogen adaptation scenarios framed by National Grid, this study outlined the importance of geospatial flexibility for both supply and demand sides, which is currently lacking in presented FES reports. The flexibility mapping shall be introduced indicating potential range of low-carbon hydrogen adaptation in various regions for each decarbonization scenario.

\subsection{FES Scenarios Comparison}

\begin{figure}[h]
\includegraphics[scale=0.445]{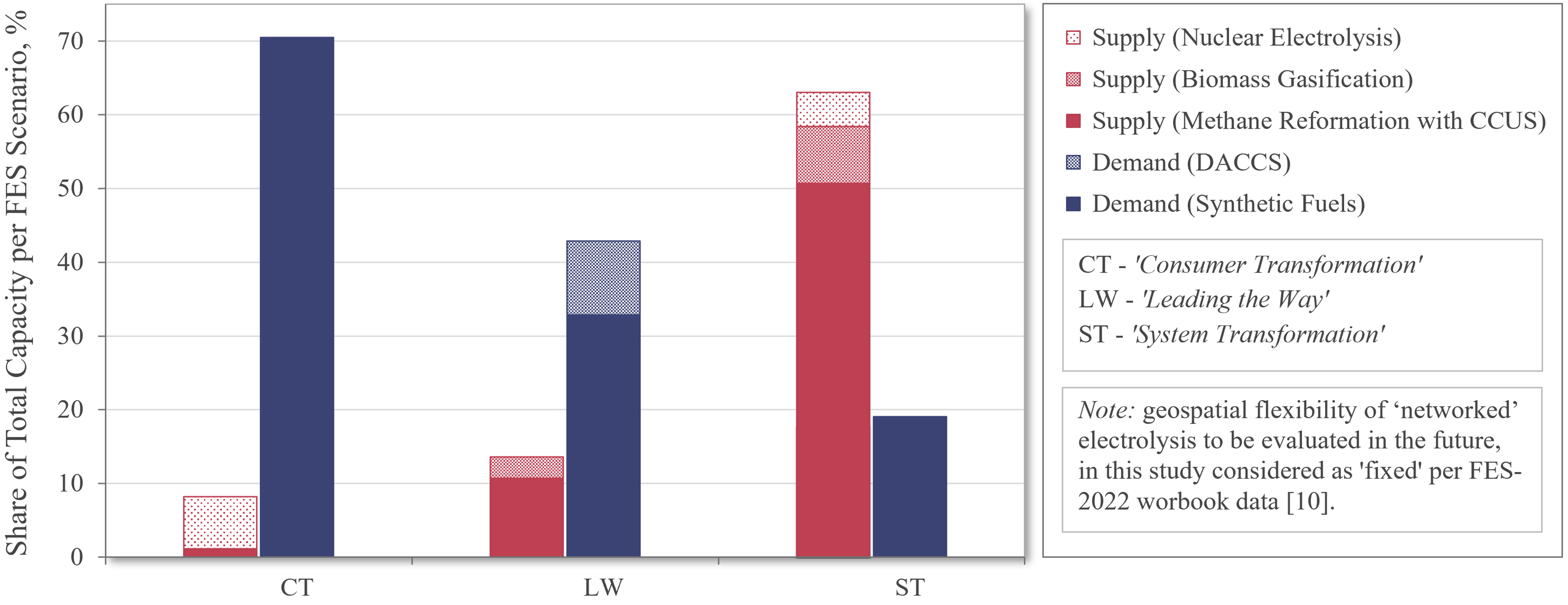}
\centering
\caption{\emph{Share of geographically 'flexible' supply and demand sectors, with capacities per FES-2022 year 2050 \cite{[8]ESO2022-Workbook}. As the share of hydrogen adaptation in GB's energy system increases, demand side flexibility decreases, enforcing a more complex hydrogen infrastructure. This is primarily due to dispersed demand for residential heating, which requires advanced infrastructure, regardless of increased supply flexibility.}} 
\label{figure6}
\end{figure}

This section discusses comparison of the FES scenarios modelled. Share of `flexible' supply and demand in relation to total low-carbon hydrogen consumption in 2050 is presented on Figure \ref{figure6}. The flexibility share shown on the Figure is based on technical aspects only, such as proximity to feedstock and essential utilities. It can be seen from Figure \ref{figure6} that share of `flexible' demand, which could potentially be located adjacent to hydrogen production, is decreasing with increased proportion of hydrogen in total energy mix while the opposite trend can be seen for supply side. In former case, the decrease in flexibility is mainly associated with adaptation of large percentage of low-carbon hydrogen in heating sector. While in latter case, the increase in supply flexibility is mainly linked to low-carbon hydrogen production through methane reformation with CCUS.

\emph{1) Hydrogen Infrastructure Scale for FES-2022 Net Zero Scenarios:}

\emph{‘Consumer Transformation’:} Local hydrogen networks with strategic location of ‘flexible’ consumers, as synthetic fuels (shipping and aviation) is deemed to serve the purpose. \emph{‘Leading the Way’:} Regional networks as not all demands can be met by local production and some residential heating is added. Location planning of synthetic fuel production (shipping and aviation) and Direct Air Carbon Capture and Storage (DACCS) will support reduction of future hydrogen infrastructure, however it is expected that regional networks would still be required to cover hydrogen distribution for residential heating. \emph{'System Transformation’:} Country scale networks, high reliance on hydrogen for heating. Overall, this scenario has the highest share of GB’s energy system reliance on hydrogen and the largest geographical area of consumers to be covered. Therefore, introduction of complex hydrogen infrastructure is essential for ‘System Transformation’ scenario.

\emph{2) Residential Heating Decarbonization Debate:}

Emissions reduction in residential heating sector is still encountering major uncertainties. While 'Consumer Transformation' scenario relies purely on domestic heating electrification, 'Leading the Way’ adopts 25\% ($\approx$42TWh) and ‘System Transformation’ 50\% ($\approx$145TWh) of hydrogen in total heating energy mix by 2050 \cite{[8]ESO2022-Workbook}. Even though heating heat pumps are by far more efficient compared to hydrogen boilers \cite{[13]ESO2022-Report}, temperature requirement cannot always be met by heat pumps particularly in apartment blocks \cite{[15]samsatli2019role}.  There is a potential for large-scale district heating sourced by heat pumps as highlighted by \citet{[60]david2017heat}, however, further research is required to justify its feasibility.

\emph{3) Hydrogen in Industry Decarbonization:} 

Review of industry sector decarbonization by BEIS \cite{[16]HMGov2023-IndustrialDecarb} shows high range of potential solutions for emissions reduction in this sector which can also be indirectly observed in FES scenarios. As example, hydrogen adaptation in  'Consumer Transformation' is less than 10\% ($\approx$11TWh) and 'System Transformation' is about 45\% ($\approx$ 87TWh) of total energy consumption by industries in 2050 \cite{[8]ESO2022-Workbook}. Even though future role of hydrogen in industry sector decarbonization has lack of clarity, the UK industrial sites are mostly concentrated in clusters which may support localised decarbonization approach reducing requirement in extensive hydrogen infrastructure.  

\emph{4) Balancing Hydrogen Storage Needs:}  

Peak hydrogen flows were observed during PyPSA-GB-H2 \cite{[63]PyPSA-GB-H2-2023} model runs in each FES scenario, which are associated with hourly temperature fluctuations and power generation peaks. These flow peaks shall see a reduction with addition of pipeline packing and increase in model resolution, however local storage of hydrogen (potentially in liquefied form) shall be considered for all scenarios to eliminate high-capacity short-term loads on infrastructure. As for geological storage, strategy for infrastructure development shall include geological storage starting from minimum hydrogen adaptation scenario and feasibility of decentralised approach.

\emph{5) GB's Net Zero Scenarios Crossroads:}

Shipping and aviation sectors are reaching the same hydrogen consumption in all FES scenarios by 2050, standing at about 70TWh and 10TWh respectively; in addition, power generation having a similar requirements in hydrogen among all FES scenarios with about 15TWh in 2050 \cite{[8]ESO2022-Workbook}. 

Based on the above, hydrogen infrastructure development shall be phased starting from localized hubs prioritizing decarbonization of power grid. This will also allow expansion of low-carbon hydrogen production sector to include ‘networked’ electrolysis. In addition, advanced planning for synthetic fuels for shipping and aviation shall be reviewed in early phase of low-carbon hydrogen hubs development. As for the existing industrial sector, it is evaluated that localized decarbonization without country-scale hydrogen infrastructure will meet the requirements. 

\subsection{Policy Recommendations}

The following areas of improvement are proposed to current policies as outcome of research completed under this study: 

(1)	Set priorities for hydrogen production hubs support under government schemes considering strategic location of future hydrogen power and synthetic fuel generation facilities. 

(2)	Further government support to research centers on residential heating, particularly multi-family house (apartment blocks). 

(3) Project Union initiated by National Gas (the UK gas system operator), to provide roadmap and phasing strategy with emphasis on self-sufficient clusters decarbonization approach.

(4)	If blending strategy is approved, ‘green light’ for hydrogen producers tapping shall be strictly monitored accounting for future strategic planning of potential final consumers. 

(5) Incorporate geospatial flexibility mapping of low-carbon hydrogen supply and demand sectors into strategic planning. The mapping shall account for techno-economic and social aspects, such as existing infrastructure, utilities, proximity to feedstock and region development including availability of transferable skills.

\subsection{Further Work}

The following aspects can be the subject of future studies of GB’s hydrogen infrastructure development:

(1)	‘Networked’ electrolysis location suggested by FES-2022 shall be re-evaluated after interconnection of PyPSA-GB-H2 \cite{[63]PyPSA-GB-H2-2023} with PyPSA-GB \cite{[11]lyden4509311open} which will allow understanding of grid constraints and economic optimization of the GB’s energy system.

(2)	Spatial resolution of PyPSA-GB-H2 model to be increased including potential location of power generation plants rather than limitation of all production and consumption to industrial cluster level. 

(3)	Pipeline packing and provision of local storage for peak loads for both power generation and heating may be incorporated in PyPSA-GB-H2 model, with re-evaluation of hydrogen power plants modelling.  

(4) Substantial difference between FES and PyPSA-GB-H2 peak hydrogen storage was observed. Peak storage requirement may affect techno-economic feasibility of low-carbon adaption and the discrepancy shall be investigated further. 

\section{Conclusions}

FES scenarios presented by National Grid \cite{[6]ESO2023-Report} on an annual basis provide potential directions of hydrogen integration in the GB's energy system, however, the extent of hydrogen application in different sectors and production capacity by each technology is highly varying between the scenarios. Moreover, FES scenarios are not providing guarantee of proceeding with one or another decarbonization scenario, in fact future energy mix will potentially land somewhere in between. Even though FES scenarios could be considered as minimum and maximum boundaries for potential hydrogen adaptation, there is a gap in definition of geospatial flexibility for each supply and demand sector. This uncertainty can lead to substantial investment in complex infrastructure which may be largely underutilized in the future. Hydrogen adaptation may have to be adjusted around the new infrastructure missing out opportunities for more efficient energy systems development. 

This study performed evaluation of GB's hydrogen infrastructure development introducing geospatial flexibility of supply and demand sectors. An open-source model PyPSA-GB-H2 was developed, improving visibility of hydrogen adaptation across various sectors and building up a base for further studies. The model incorporated all potential consumers and producers of low-carbon hydrogen predicted in FES scenarios and is flexible to be re-used along with future developments in this sector. A general nature of PyPSA-GB-H2 and variety of covered sectors will also allow to adopt the model to other European Union (EU) countries, modifying GIS location and low-carbon hydrogen capacities in line with specific country strategic planning. 

Development of hydrogen infrastructure and hydrogen economy are two interdependent aspects. From one side, the scale of infrastructure shall be defined by hydrogen integration in the future energy mix. From another, hydrogen supply chain growth may be constrained by the absence of facilities to link potential supply and demand sites.

This study recommends prioritising the establishment of green hydrogen hubs, in the near-term, aligning with demands for synthetic fuels production, industry, and power, which can facilitate the subsequent roll out of up to 10GW of hydrogen production capacity by 2050. Additionally, the analysis quantifies a high proportion of hydrogen supply and demand which can be located flexibly.

\section{Supplementary Material}

More details on PyPSA-GB-H2 modelling approach can be found on a public GitHub repository -  \href{https://github.com/tatyanadergunova/PyPSA-GB-H2.git}{https://github.com/tatyanadergunova/PyPSA-GB-H2.git}.



\bibliographystyle{model1-num-names}
\bibliography{elsarticle-template-harv.bib}





\end{document}